\documentclass[a4paper,12pt]{article} 
\usepackage{authblk}
\usepackage{blindtext}
\usepackage[english]{babel}
\usepackage{float}
\usepackage{graphicx}
\usepackage{soul}

\title{Comment on: \\
Lipp, J.,  Banerjee, R., Patwary, MF., Patra, N., Dong, AH., Girgsdies, F., Bare, SR. \& Regalbuto, JR.;
'Extension of Rietveld Refinement for Benchtop Powder XRD Analysis of Ultrasmall Supported Nanoparticles', Chem. Mater. 2022, 34, 18, 8091-8111.}

\begin{document}                  

\title{Comment on: \\
Lipp, J.,  Banerjee, R., Patwary, MF., Patra, N., Dong, AH., Girgsdies, F., Bare, SR. \& Regalbuto, JR.;
'Extension of Rietveld Refinement for Benchtop Powder XRD Analysis of Ultrasmall Supported Nanoparticles', Chem. Mater. 2022, 34, 18, 8091-8111.}

\author[1]{Zbigniew Kaszkur}

\affil[1]{Institute of Physical Chemistry PAS,Kasprzaka 44/52, 01-224, Warszawa, Poland}

\maketitle                        





\section{Comments}
The article deals with application of Rietveld refinement to multiphase powder containing ultrasmall nanoparticles. A number of the described methods is sound and the presented results looks reasonably, however the claims given in the title and some conclusions have to be wrong for principal reasons. The Rietveld method describes analytically each participating crystal phase and treats the measured pattern as a sum of the phases patterns. Another fundamental approach to powder diffraction is  Debye`s scattering equation where the summation is done not over crystal phases but over individual waves scattered on pairs of atoms, averaged over all possible orientations of the interatomic distance vector versus the scattering vector (as for perfect powder). The difference between these two methods is fundamental. If we assume that our sample consists of two phases, one composed of atoms A and the second, composed of atoms B, the Rietveld approach considers only peaks of two crystal phases described only by distances A-A and B-B. On the other hand, Debye`s scattering equation includes all distances A-A, B-B and A-B. For the two phases remaining in close, epitaxial contact, the A-B term cannot be neglected, the difference between the two approaches becomes meaningful and the Debye's approach is correct but the Rietveld's is wrong. 

The figure below, as an example, shows diffraction pattern of a Pt nanoparticle consisting of 561 atoms calculated via Debye`s scattering equation as for Cu $K_{\alpha}$ radiation. From the original shape of cubooctahedron of ~2 nm size, the surface atoms (marked as SRF) were expanded via energy relaxation to conform to SRF-SRF interatomic distance of about 0.316 nm (like Pt-Pt in $PtO_{2}$ phase). The final model was supposed to closely resemble a core-shell particle with the core - Pt fcc nanocrystal (1.5 nm size), surrounded by the shell (0.6 nm thick) resembling $PtO_{2}$. The model neglects scatering by oxygen atoms in $PtO_{2}$, but considering number of electrons in oxygen atom (8) with the number of electrons in Pt atom (78), one has to admit that it is quite realistic approximation. The important message from the figure is that the contribution from core-SRF distances is meaningful and it has complex form with negative values changing character of the pattern. For smaller nanocrystals, this contribution becomes more important. It cannot be neglected and any analysis of the overall pattern (black curve) in terms of two phase contributions (like Rietveld`s) to intensity is wrong. All the calculations illustrating the above conclusions can be easily repeated using developed by us software Cluster freely available to download \cite{cluster}. It contains model builder, a number of simulation technics and uses interatomic n-body potential model after Sutton\&Chen \cite{Sutton}, known to perform well for metals. However the estimate of the core-SRF contribution does not depend strongly on the potential model.

\begin{figure}
 \centering
  \includegraphics[width=1\columnwidth]{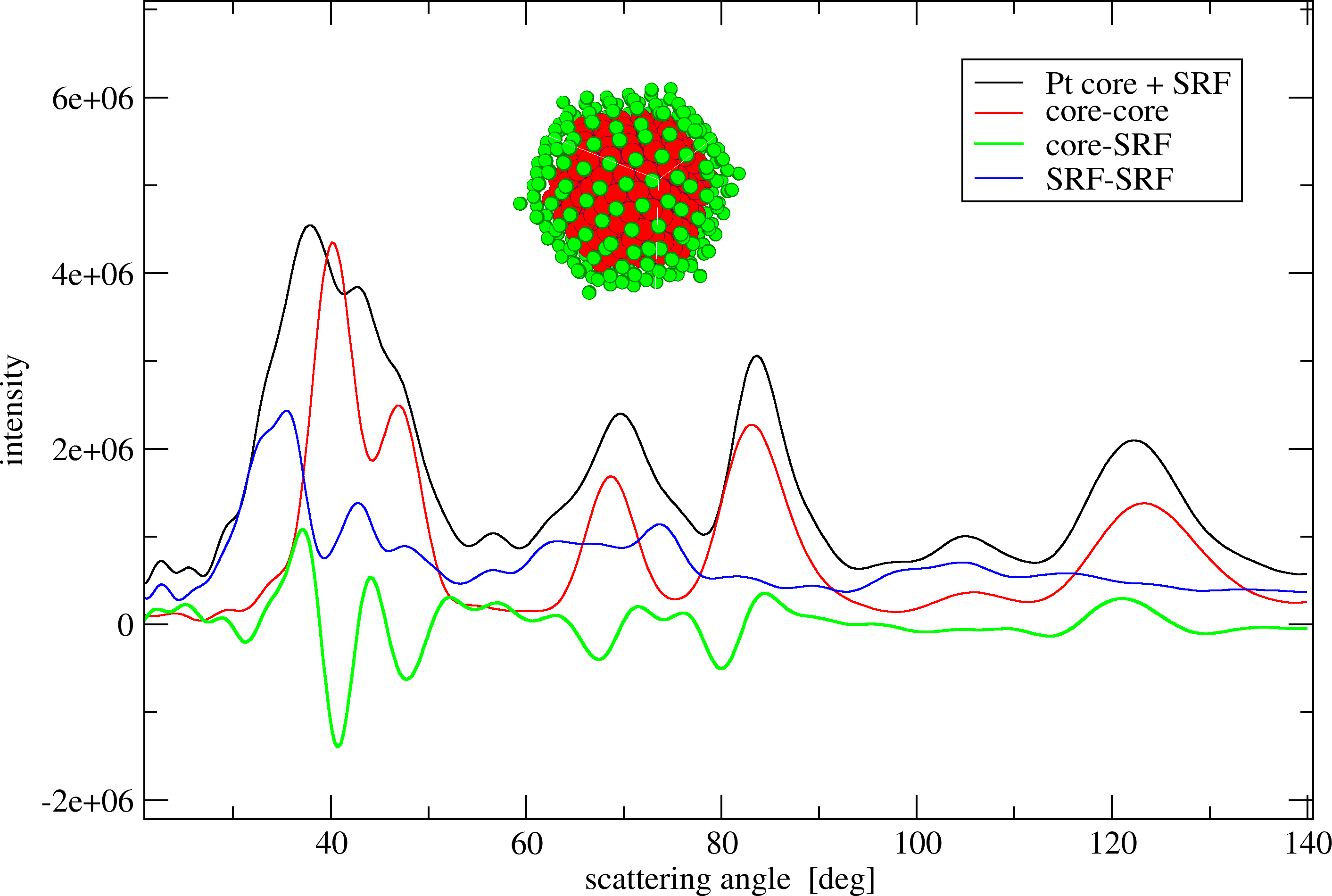}
 \caption{Diffraction pattern (Cu $K_{\alpha}$ radiation) of $Pt-PtO_{2}$ core-shell(SRF) nanocrystal (black line) with contribution from core-core distances (red line), SRF-SRF distances (blue line) and core-SRF distances (green line).}

\end{figure}

The conclusion of the authors that they observe core-shell system of Pt and $PtO_{2}$ has to be wrong. The most likely the structure contains separate nanocrystals of Pt (with chemisorbed oxygen) and of  $PtO_{2}$.

The most important conclusion is that the Rietweld refinement method in most cases cannot be applied to complex nanocrystalline structures and the reasons are of fundamental character. There is a number of other arguments suggesting that for nanocrystals (e.g. metals) of size less than 10 nm, many interesting structural features are neglected in the Rietveld analysis tending to smooth out deviations from the perfect crystallinity. One of them is the peaks shifting from the Bragg`s position described by the authors as peak pull. There is a number of factors causing such shifts including diffractometer errors, multiplying the pattern by the sloped atomic factors, Lorentz-polarization factor etc. but at the bottom of it remains the intrinsic peak shift connected with nanocrystal surface relaxation (e.g. contraction) that is usually different for different crystal faces. The shift can then differ for various peaks, even in direction. The experimental monitoring of such shifts in situ can provide valuable information on the state of the surface. Technics exploiting these phenomena are being developed by us over the last 20 years and were reported \cite{Kaszkur:nb5191,Kaszkur:he5654,B820510H}.




     



\end{document}